# Target decoupling in a coupled optical system resistant to random perturbation

Sunkyu Yu, Xianji Piao, and Namkyoo Park*

*Photonic Systems Laboratory, Department of Electrical and Computer Engineering, Seoul National University, Seoul 08826, Korea*

*E-mail address for correspondence: nkpark@snu.ac.kr

**To suppress unwanted crosstalks between nearby optical elements, the decoupling technique for integrated systems has been desired for the target control of light flows. Although cloaking methods have enabled complete decoupling of optical elements by manipulating electromagnetic waves microscopically, it is neither feasible nor necessary to control each unit element in coupled systems when considering severe restrictions on material parameters for cloaking. Here we develop the macroscopic approach to design crosstalk-free regions in coupled optical systems. By inversely designing the eigenstate which encompasses target elements, the stable decoupling of the elements from the coupled system is achieved, being completely independent from the random alteration of the decoupled region, and at the same time, allowing coherent and scattering-free wave transport with desired spatial profiles. We also demonstrate the decoupling in disordered systems, overcoming the transport blockade from Anderson localization. Our results provide an attractive solution for 'target hiding' of elements inside coupled systems.**

Invisibility cloaking is one of the most fascinating achievements in transformation optics[1-3]. The coordinate transformation between virtual and physical spaces provides the rigorous design guidance of material parameters, perfectly separating the light flow in the cloaked region from that in the other part. Although transformation optics derived from full-vectorial Maxwell's equations[1] successfully provides an exact solution for omnidirectional and scattering-free perfect cloaking, at the same time, its strict demand on material designs has caused hardship to the practical implementation of the cloaking in spite of recent achievements in optical metamaterials[4].

The stringent condition of rigorous transformation optics has also hindered the application of the cloaking to photonic integrated circuits which require the "decoupling" technique[5,6] between elements for crosstalk-free signal transport. Consider the 'hiding' (or 'decoupling') of some elements inside densely packed coupled optical systems[5,7-11]. Transformation optics in this scenario provides the severely intricate solution even for the approximated case[12]: the coating of target elements with spatially-varying, highly anisotropic metamaterials of extreme material parameters (effective permittivity ~ 0), which derives the 'microscopic' removal of the coupling to the target elements. We note that similar restrictions can also be found in other alternative cloaking methodologies. The cloaking using accidental degeneracy[13] requires the well-defined crystalline structure to maintain the Dirac point, and thus cloaked elements should be separated by more than several lattice periods, prohibiting the integration. Although the concept of parity-time symmetry has been applied to the unidirectional invisibility in one-dimensional coupled structures[14,15] based on their singular scattering, the extension to multi-dimensional integrated systems encounters the similar difficulty with transformation optics: the coating of spatially varying gain-loss media[16] for each element. The optical analogy of the adiabatic passage[5,17] has also been employed to hide the inner waveguide in tri-atomic designs, but its multi-dimensional or *N*-atomic realization still remains as a challenge.

Here, we propose the 'macroscopic' approach to the decoupling based on the eigenstate molding applicable to *N*-atomic coupled optical systems, instead of the microscopic material arrangement for each element[1-3,13,16]. We demonstrate that the scattering-free perfect transmission can be achieved through the system eigenstate which includes target decoupled elements, against the random perturbation of the self-energy inside the target region of the system. By controlling the self-energy of the system in a moderate range, the designer spatial profile of the wave flow can also be achieved around target elements, while preserving the scattering-free condition. Utilizing the generality of our eigenstate decoupling method, we also show the stable decoupling in disordered systems for the first time, which resolves the blockade of wave transport from Anderson localizations[18,19].

We begin with an instructive example of a triatomic system where each element has the self-energy of $\rho_i$ (*e.g.* resonant frequency $f$ of an uncoupled resonator), and the coupling between the $i$-th and $j$-th elements is given as $\kappa_{ij}$ (Fig. 1a, $\kappa_{ij} \sim \kappa_{ji}$ for the similar shape of elements[20]). The system then satisfies the following Hamiltonian equation[5,10,21]

$$\begin{bmatrix} \rho_1 & \kappa_{12} & \kappa_{13} \\ \kappa_{21} & \rho_2 & \kappa_{23} \\ \kappa_{31} & \kappa_{32} & \rho_3 \end{bmatrix} \begin{bmatrix} \psi_1 \\ \psi_2 \\ \psi_3 \end{bmatrix} = \rho \begin{bmatrix} \psi_1 \\ \psi_2 \\ \psi_3 \end{bmatrix}, \qquad (1)$$

for the field amplitude at each element $\Psi = [\psi_1, \psi_2, \psi_3]^T$. We establish the decoupling of the 3$^{rd}$ element, calling for the invariant eigenstate for the random perturbation of $\rho_3$ (Fig. 1a *vs* 1b, as $\rho_{3a} \neq \rho_{3b}$). From the setting of $\psi_3 = 0$ to remove the $\rho_3$-dependency, *i.e.* 'hiding' of the 3$^{rd}$ element in the target eigenstate, Eq. (1) then derives the condition of $\kappa_{31} \cdot \psi_1 + \kappa_{32} \cdot \psi_2 = 0$ which corresponds to the destructive coupling interference in the 3$^{rd}$ element (Fig. 1a,b). This condition applied to Eq. (1) defines the necessary condition of the self-energy for decoupling the 3$^{rd}$ element as $\rho_1 - \rho_2 = \kappa_{31} \cdot \kappa_{12}/\kappa_{32} - \kappa_{32} \cdot \kappa_{21}/\kappa_{31}$, and the corresponding eigenvalue of the target eigenstate can be controlled by $\rho = \rho_1 - \kappa_{31} \cdot \kappa_{12}/\kappa_{32}$. Hence, by controlling the self-energy of the elements ($\rho_{1,2}$) which have the given coupling network (fixed $\kappa_{ij}$), we can 'hide' some elements inside the coupled system at the desired eigenvalue $\rho$, for any networks even including irregular or symmetry-broken cases (*e.g.* $\kappa_{23} \neq \kappa_{31}$). We note that this approach can be easily extended to hiding *m*-elements inside *N*-atomic systems (Fig. 1c, Supplementary Note 1). Interestingly, although the nearby elements (blue and red elements in Fig. 1c) of the target region (2 dark gray elements in the center, Fig. 1c) should have the designed field distribution for the decoupling, the field at the rest elements (light gray elements in Fig. 1c) of the system can be controlled irrespective of the decoupling (Supplementary Note 1 and Fig. S1c,d), allowing the scattering-free designer wave flow around the decoupled region.

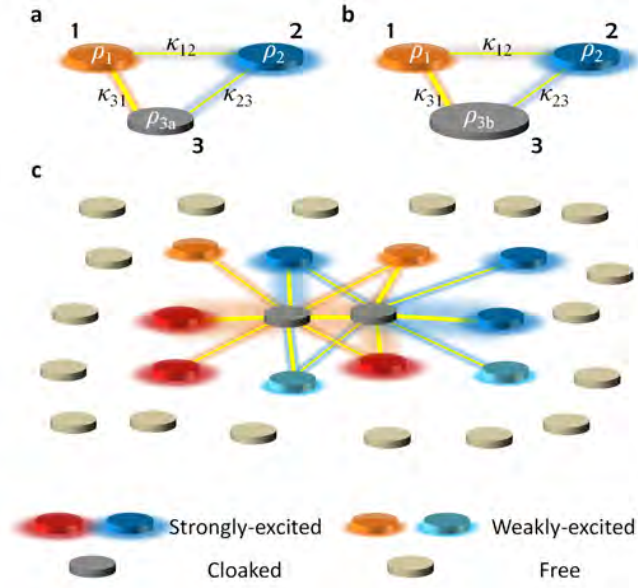

**Figure 1. Schematics of the designer state for the decoupling in coupled optical systems.** Tri-atomic examples for different self-energy at the 3$^{rd}$ element: (**a**) $\rho_{3a}$ and (**b**) $\rho_{3a}$. $\rho$ for self-energy of each element and $\kappa$ for the coupling between elements in (**a,b**). (**c**) *N*-atomic example for two target decoupled elements at the center (dark gray). Coupling is denoted as the line between elements, and for clarity, coupling terms only around the target elements are presented.

Based on the design methodology in Supplementary Note 1, we demonstrate the decoupling in coupled optical systems (Fig. 2). Without loss of generality, we employ the system of coupled titanium oxide ($TiO_2$) circular resonators embedded in an indium antimonide (InSb) crystalline compound, operating in the terahertz regime with transverse magnetic (TM) monopole resonances. We control the radii of resonators and their location to adjust the resonant frequency *f* and coupling $\kappa$, respectively (see the detailed design in Supplementary Note 2). We investigate the 11 × 11 coupled resonator square lattice, encompassing the 3 × 3 decoupled region at the center of the system (the 'decoupled region D in Fig. 2. Its surrounding 'transport' region is denoted as T). The binary random self-energy is applied to the resonators in the region D for clarity; the elements inside the decoupled region have one of the two self-energy values (or resonant frequencies) $f = f_0$ or $f = 1.1 \cdot f_0$ with the same probability ($f_0$: operating frequency). The self-energy distribution of the region T is derived both for decoupling and designed spatial profiles of wave transport, following Supplementary Note 1. To demonstrate the decoupling operation, we compare the results from the eigenstate decoupling environments (Fig. 2b,e,h) with those from the ordinary crystal environments which have identical elements at the region T (Fig. 2c,f,i).

Figures 2a-c and 2d-f show the wave transfer for the different sets of elements inside the target region D. In general, the detailed configuration of the self-energy distribution strongly affects the wave transport in the coupled optical system, because the self-energy determines not only the phase evolution inside each element but also the coupling efficiency between elements[22]. However, regardless of the configuration of the target region D (D ≠ D' in Fig. 2a,d), the eigenstate decoupling systems provide the perfect plane wave transfer (Fig. 2b,e) with the same transport region configuration (same T in Fig. 2a,d), in sharp contrast to strong scattering and spatial incoherence in the crystal platforms the light flow of which has also strong dependence on the configuration of the region D (D ≠ D' in Fig. 2c,f). This result demonstrates that the decoupling eigenstate designed by the methodology in Note S1 successfully neglects the self-energy perturbation inside the target region, realizing the "target decoupling" based on the form of the eigenstate.

As shown in the closed form of Eq. (S5) in Supplementary Note 1, the self-energy distribution is uniquely defined for any nodeless eigenstate which satisfies the decoupling condition ($\psi = 0$) in the region D. Conversely, by controlling the self-energy of the environmental region T (T' in Fig. 2g), the molding of the spatial form of wave flows becomes possible while preserving the scattering-free condition around the region D; as shown in the wave focusing example in Fig. 2h (compared to the random scattering in the ordinary environment of Fig. 2i). We thus note that designer wave flows with optical functionalities, such as focusing, beam splitting, and mode conversion, can be achieved, regardless of the perturbation inside the target decoupled region D.

Although the examples of Fig. 2a-i are based on the lattice structure, the main strength of the eigenstate decoupling is the high applicability to 'any' coupling networks, in contrast to the indispensable spatial symmetry in the Dirac point cloaking[13] or parity-time-symmetric invisibility[14-16]. The evidence is shown in Fig. 2j-2l, demonstrating the decoupling in the system which has the off-diagonal disorder[23,24] from the random deformation of each resonator position (disordered coupling both in $D_d$ and $T_d$ regions in Fig. 2j). Perfect coherent transmission (Fig. 2k) is achieved as same as the cases in the lattice structure, overcoming the incoherent blockade of wave transport from Anderson localization (35dB enhancement from 0.03% transmission at Fig. 2l). Distinct from previous cloaking methods[13-16] which necessitate the strict spatial symmetry for the position of each optical element, the eigenstate decoupling method allows for the decoupling inside randomly distributed resonator systems, surprisingly, compensating the Anderson blockade from the off-diagonal disorder, as an example of the designer disorder[25-29].

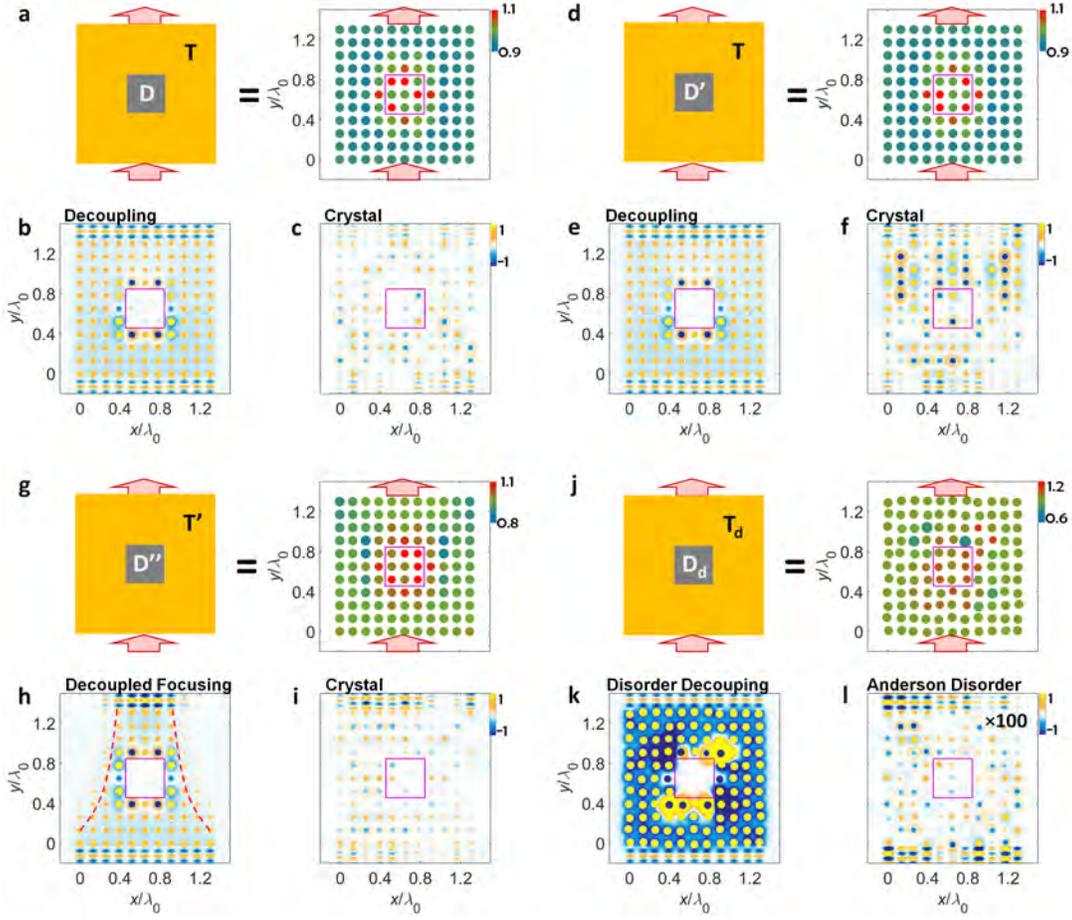

**Figure 2. Demonstration of eigenstate decoupling.** The decoupling with plane wave input and output is shown in (**a,b**) and (**d,e**) compared to the cases of ordinary crystal systems with identical elements in (**c,f**), for the different configuration in the decoupled region (D ≠ D', red boxes in (**a,d**)). The decoupling with wave focusing (T') is shown in (**g,h**) compared to the case of the ordinary background in (**i**). The decoupling in the disordered system ($D_d$, $T_d$) is shown in (**j,k**) compared to the case of ordinary Anderson off-diagonal disorder in (**l**). The position of each resonator in (**j-l**) is randomly deformed for *x* and *y* axes, with the ±$Λ_0$/10 maximum deformation for the original periodicity $Λ_0$. The field amplitude in (**l**) is magnified (×100) for the presentation. $λ_0$ is the free-space wavelength, and all of the design parameters are shown in Supplementary Note 2.

Illustrating the stability and spectral property of the eigenstate decoupling method, the statistical spectral analysis of the decoupling system (Fig. 2a-f) is also shown in Fig. 3. For 9 decoupled elements which have binary random resonant frequencies of $f = f_0$ and $f = 1.1 \cdot f_0$, the statistical ensemble of $2^9$ samples is realized to examine the coherence and transmission over the decoupling system (each thin lines in Fig. 3a,b). About 94% of average transmission (Fig. 3a) with the uniform spatial profile (Fig. 3b) is achieved near the operating frequency, robust to the random alteration of the decoupled region (~0.040% standard deviation for the transmission): in sharp contrast to the performance of the ordinary crystal system (~16% transmission with 11% standard deviation).

The output flow through the decoupling system preserves excellent spatial coherence as well. Compared to incoherent scattering with random phase and amplitude in the ordinary crystal system (Fig. 3d,f), the decoupling system of Fig. 2a derives the unity amplitude (Fig. 3c) and constant phase (Fig. 3e) at the output, independent from the random alteration of the decoupled region. The proposed eigenstate decoupling system thus preserves all of the spatial information of the incident wave regardless of the detailed composition of the decoupled region, realizing the complete decoupling condition.

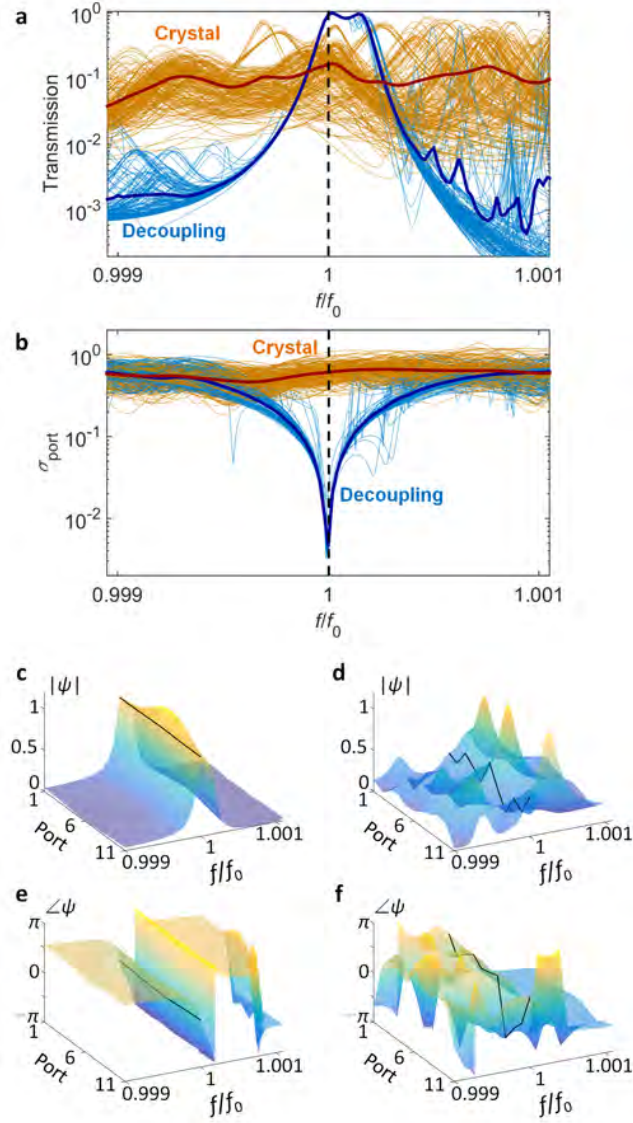

**Figure 3. Statistical spectral analysis of eigenstate decoupling.** (**a**) Transmission and (**b**) amplitude fluctuation spectra for the decoupling system (light blue thin lines) and the ordinary crystal system (orange thin lines), for the ensemble of $2^9$ samples. The fluctuation $\sigma_{port}$ in (**b**) is the standard deviation of output field amplitude for 11 ports, normalized by the averaged amplitude ($\sigma_{port} = 0$ for ideal plane wave). Blue and red thick lines in (**a,b**) denote the averaged results for $2^9$ samples of each system. Black dashed line depicts the design frequency. The amplitude (**c,d**) and phase (**e,f**) of the output field is plotted as a function of frequency and output positions, for an example of decoupling (**c,e**) and ordinary systems (**d,f**). Black lines denote the results at the operating frequency $f_0$.

In summary, we proposed a new class of decoupling techniques for photonic integrated circuits, the macroscopic 'decoupling' of optical elements, by exploiting the system eigenstate with destructive interference regions. Based on the statistical analysis, we proved that the eigenstate decoupling method stably hides optical elements inside the coupled system, simultaneously allowing coherent wave transport with desired spatial profiles. Distinct from previous achievements in symmetry-based cloaking[13-16], we also demonstrated the decoupling in disordered systems with the suppressed Anderson localization, as an example of the designer disorder[25-29].

The eigenstate decoupling method provides excellent flexibility to the waveform molding in coupled optical systems, with the control of transport region elements. Likewise the global scattering increase in spectral domain as observed in most of cloaking structures[30] (except few extreme cases such as diamagnetic and superconducting cloaks[30]), the bandwidth problem in our system is the engineering subject which can be improved by alleviating the strict decoupling condition. Our approach, separating target elements from the other region in coupling networks using moderate material/structural parameters, also possesses the link with the selective target control[31-33] in network theory.

**Acknowledgments**

This work was supported by the National Research Foundation of Korea (NRF) through the Global Frontier Program (GFP, 2014M3A6B3063708), the Global Research Laboratory Program (GRL, K20815000003), and Korea Research Fellowship Program (KRF, 2016), which are all funded by the Ministry of Science, ICT & Future Planning of the Korean government. S. Yu was also supported by the Basic Science Research Program (2016R1A6A3A04009723) through the NRF, funded by the Ministry of Education of the Korean government.

**Author Contributions**

S.Y. conceived the presented idea. S.Y. and X.P. developed the theory and performed the computations. N.P. encouraged S.Y. to investigate the inverse design of the eigenstate for decoupling structures while supervising the findings of this work. All authors discussed the results and contributed to the final manuscript.

**Competing Interests Statement**

The authors declare that they have no competing financial interests.

**Supplementary Note 1. Design of an *N*-atomic coupled system for a decoupling eigenstate**

Extending the discussion in the main text, here we show the design procedure (Fig. S1) of the *N*-atomic system which possesses a decoupling eigenstate. The Hamiltonian equation for the *N*-atomic system composed of weakly coupled elements is[1-4]

$$\begin{bmatrix} \rho_1 & \kappa_{12} & \cdots & \kappa_{1N} \\ \kappa_{21} & \rho_2 & \cdots & \kappa_{2N} \\ \vdots & \vdots & \ddots & \vdots \\ \kappa_{N1} & \kappa_{N2} & \cdots & \rho_N \end{bmatrix} \begin{bmatrix} \psi_1 \\ \psi_2 \\ \vdots \\ \psi_N \end{bmatrix} = \rho \cdot \begin{bmatrix} \psi_1 \\ \psi_2 \\ \vdots \\ \psi_N \end{bmatrix}. \tag{S1}$$

For the given coupling network ($\kappa_{jk}$, Fig. S1a), we will derive the necessary form of the decoupling eigenstate $\Psi = [\psi_1, \psi_2, \ldots, \psi_N]^T$, to determine the corresponding self-energy[4] $\Omega = [\rho_1, \rho_2, \ldots, \rho_N]^T$.

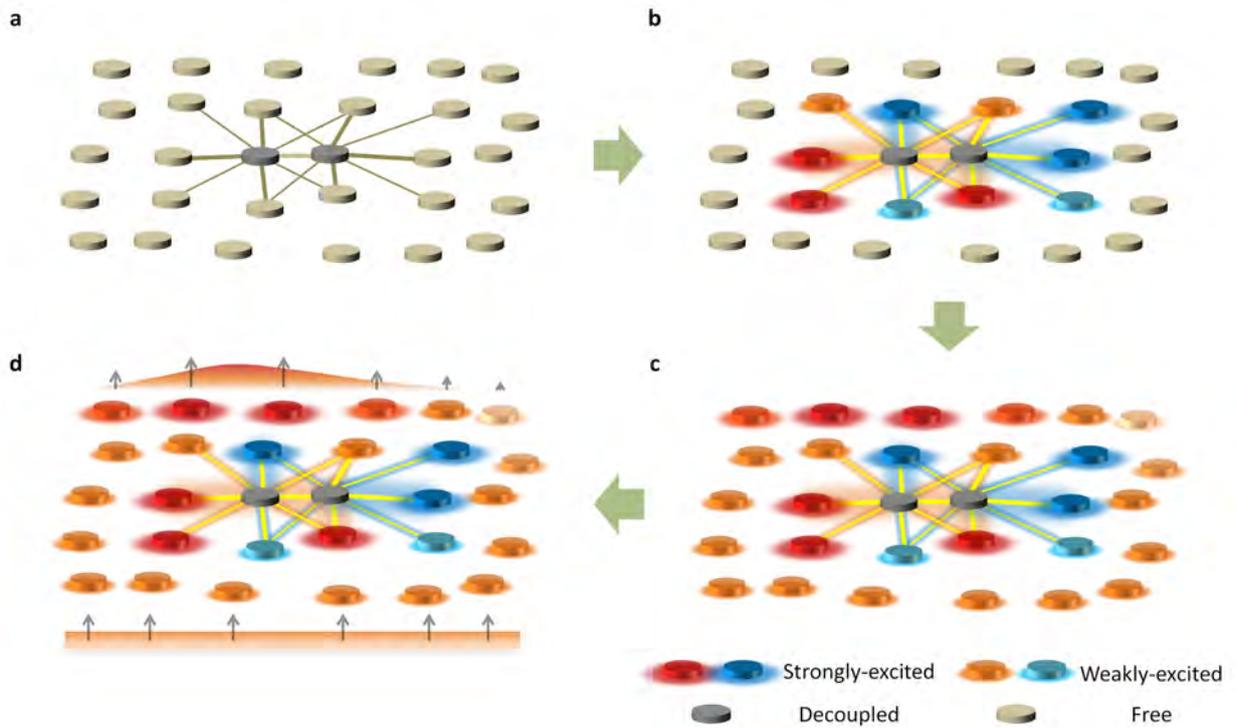

**Figure S1. The design procedure of an *N*-atomic system for a decoupling eigenstate.** (**a**) The selection of target decoupled elements in the system which has the determined coupling network. The design of the field distribution in (**b**) nearby elements of target elements and (**c**) the rest tunable elements. (**d**) The wave flow through the system, realizing the scattering-free manipulation of the waveform. Couplings only around the decoupled elements are presented for clarity.

With $m$ number of decoupled elements ($m \leq N$, $\psi = 0$) the indices of which constitute the set A, the set of nearby elements ($\psi \neq 0$) for each decoupled element can be defined as $B_j$ ($j \in$ A, e.g. A = {5} and $B_5$ = {1,2,3,4,6,7,8,9} in Fig. S2a, for $m = 1$ and $N = 9$). Because at most nearest-neighbor and next-nearest-neighbor coupling coefficients are significant in realistic structures[5] (Fig. S1a) due to the exponential decay of evanescent coupling in space, each row of Eq. (S1) for decoupled elements derives the following condition of the destructive coupling interference as

$$\sum_{k \in B_j} \kappa_{jk} \cdot \psi_k = 0, \quad (S2)$$

where $j \in$ A, and the condition of $k \in B_j$ represents the nearby coupling ($\kappa_{jk} \sim 0$ for far-off elements of $k \notin B_j$, for the $j$-th decoupled element). Equation (S2) governs the necessary condition of the field amplitude in nearby elements (Fig. S1b). Note that the degree of freedom (DOF) of Eq. (S2) is determined by the number of nearby elements for each decoupled element (Fig. S2).

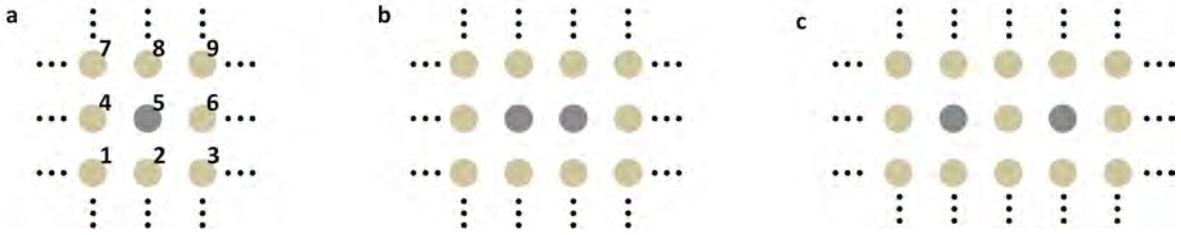

**Figure S2. The schematics of decoupled (dark gray) and nearby (light gray) elements**: (**a**) 7 DOF, (**b**) 8 DOF, and (**c**) 11 DOF.

After the nearby elements for all of the decoupled elements are determined to satisfy Eq. (S2) (Fig. S1b, the elements in $B_j$ for all $j \in$ A), the other region of the decoupling eigenstate can then be designed (Fig. S1c). Except the rows of decoupled indices for Eq. (S2), the other part of Eq. (S1) has the form of

$$\begin{bmatrix} \rho_{s-1} & \kappa_{s-1,s-2} & \cdots & \kappa_{s-1,s-(N-m)} \\ \kappa_{s-2,s-1} & \rho_{s-2} & \cdots & \kappa_{s-2,s-(N-m)} \\ \vdots & \vdots & \ddots & \vdots \\ \kappa_{s-(N-m),s-1} & \kappa_{s-(N-m),s-2} & \cdots & \rho_{s-(N-m)} \end{bmatrix} \begin{bmatrix} \psi_{s-1} \\ \psi_{s-2} \\ \vdots \\ \psi_{s-(N-m)} \end{bmatrix} = \rho \cdot \begin{bmatrix} \psi_{s-1} \\ \psi_{s-2} \\ \vdots \\ \psi_{s-(N-m)} \end{bmatrix}, \quad (S3)$$

where the new index (s-p) $\notin$ A and $1 \leq$ (s-p) $\leq N$. For the subset of the decoupling eigenstate $\Psi_s$ =

$[\psi_{s\text{-}1}, \psi_{s\text{-}2},\ldots, \psi_{s\text{-}(N\text{-}m)}]^T$, although the nearby elements of decoupled elements ($(s\text{-}p) \in B_j$ for all $j \in A$, red and blue elements in Fig. S1b) are already determined for the decoupling (Eq. (S2)), the other elements ($(s\text{-}p) \notin B_j$ for any $j \in A$, light gray elements in Fig. S1c) can be freely set to achieve desired optical functionalities (e.g. steered beam focusing in Fig. S1d), finally defining $\Psi_s$ and then $\Psi$ where $\psi_j = 0$ for $j \in A$.

From the decoupling eigenstate $\Psi$ with the desired functionality, we can then achieve the corresponding self-energy distribution $\Omega = [\rho_1, \rho_2,\ldots, \rho_N]^T$. The $(N-m) \times (N-m)$ matrix equation of Eq. (S3), off-diagonal terms of which have given values, can be recast into the form of

$$\begin{bmatrix} \psi_{s\text{-}1} & 0 & \cdots & 0 \\ 0 & \psi_{s\text{-}2} & \cdots & 0 \\ \vdots & \vdots & \ddots & \vdots \\ 0 & 0 & \cdots & \psi_{s\text{-}(N-m)} \end{bmatrix} \begin{bmatrix} \rho_{s\text{-}1} \\ \rho_{s\text{-}2} \\ \vdots \\ \rho_{s\text{-}(N-m)} \end{bmatrix} = \left( \rho \cdot I - \begin{bmatrix} 0 & \kappa_{s\text{-}1,s\text{-}2} & \cdots & \kappa_{s\text{-}1,s\text{-}(N-m)} \\ \kappa_{s\text{-}2,s\text{-}1} & 0 & \cdots & \kappa_{s\text{-}2,s\text{-}(N-m)} \\ \vdots & \vdots & \ddots & \vdots \\ \kappa_{s\text{-}(N-m),s\text{-}1} & \kappa_{s\text{-}(N-m),s\text{-}2} & \cdots & 0 \end{bmatrix} \right) \begin{bmatrix} \psi_{s\text{-}1} \\ \psi_{s\text{-}2} \\ \vdots \\ \psi_{s\text{-}(N-m)} \end{bmatrix}, \text{(S4)}$$

where $I$ is the $(N-m) \times (N-m)$ identity matrix. Because the diagonal matrix $diag(\Psi_s)$ has its inverse due to $\psi_{(s\text{-}p)} \neq 0$ for all $(s\text{-}p) \notin A$, Eq. (S4) derives the required self-energy distribution $\Omega_s = [\rho_{s\text{-}1}, \rho_{s\text{-}2},\ldots, \rho_{s\text{-}(N\text{-}m)}]^T$ except the decoupled elements as,

$$\begin{bmatrix} \rho_{s\text{-}1} \\ \rho_{s\text{-}2} \\ \vdots \\ \rho_{s\text{-}(N-m)} \end{bmatrix} = \begin{bmatrix} \psi_{s\text{-}1} & 0 & \cdots & 0 \\ 0 & \psi_{s\text{-}2} & \cdots & 0 \\ \vdots & \vdots & \ddots & \vdots \\ 0 & 0 & \cdots & \psi_{s\text{-}(N-m)} \end{bmatrix}^{-1} \left( \rho \cdot I - \begin{bmatrix} 0 & \kappa_{s\text{-}1,s\text{-}2} & \cdots & \kappa_{s\text{-}1,s\text{-}(N-m)} \\ \kappa_{s\text{-}2,s\text{-}1} & 0 & \cdots & \kappa_{s\text{-}2,s\text{-}(N-m)} \\ \vdots & \vdots & \ddots & \vdots \\ \kappa_{s\text{-}(N-m),s\text{-}1} & \kappa_{s\text{-}(N-m),s\text{-}2} & \cdots & 0 \end{bmatrix} \right) \begin{bmatrix} \psi_{s\text{-}1} \\ \psi_{s\text{-}2} \\ \vdots \\ \psi_{s\text{-}(N-m)} \end{bmatrix}. \text{(S5)}$$

Because the satisfaction of Eq. (S2) and Eq. (S3) corresponds to the satisfaction of Eq. (S1), the self-energy distribution $\Omega$ which has the subset of $\Omega_s$ from Eq. (S5) and 'arbitrary' values for the $\Omega_s$'s complementary set, derives the decoupling eigenstate $\Psi$ which has $\psi_j = 0$ for $j \in A$ and $\Psi_s$ for the other part. We note that the eigenstate $\Psi$ in the potential $\Omega$ therefore achieves the decoupling (scattering-free for arbitrary $\rho_j$ of $j \in A$) and the functionality (designed $\Psi_s$, Fig. S1d) at the same time.

**Supplementary Note 2. Design of transverse magnetic monopole resonances**

In the main text, we utilize the transverse magnetic 'monopole' mode ($H_z$ and $E_{r,\varphi}$ fields with $\partial H_z / \partial \varphi = 0$) of two-dimensional circular resonators (the refractive index $n = n_1$ for dielectric core and $n = -in_2$ for metallic background), which derives the coupling coefficient dependent only on the distance between resonators. From the governing wave equation

$$r \cdot \frac{\partial^2 H_z}{\partial r^2} + \frac{\partial H_z}{\partial r} + k_0^2 n^2 r \cdot H_z = 0, \quad (S6)$$

where $k_0 = 2\pi \cdot f / c$ is the free-space wavenumber, we seek the localized solution of $H_z$ without singularity, which has the form of

$$\begin{aligned} H_z &= c_1 \cdot J_0(n_1 k_0 r) & (\text{for } r \leq r_0) \\ &= c_2 \cdot K_0(n_2 k_0 r) & (\text{for } r > r_0) \end{aligned}, \quad (S7)$$

where $J_0$ and $K_0$ each denotes the zeroth-order Bessel and modified Bessel function. Equation (S7) then derives the resonance condition of

$$n_1 \cdot J_0(n_1 k_0 r_0) \cdot K_1(n_2 k_0 r_0) + n_2 \cdot J_1(n_1 k_0 r_0) \cdot K_0(n_2 k_0 r_0) = 0, \quad (S8)$$

from the electromagnetic boundary condition ($H_{z1} = H_{z2}$ and $E_{\varphi 1} = E_{\varphi 2}$).

We assume a titanium oxide[6,7] core (TiO$_2$, refractive index $n = 10$) and an indium antimonide crystalline compound[8,9] background (InSb, $n = 0.3619 - 5.107i$), for the operation in the terahertz regime (near 1.1 THz). Equation (S8) then has a solution for $k_0 \cdot r_0 = 0.275$, which derives the necessary core radius for the resonant frequency $f_0$ as

$$r_0 = 0.275 \cdot \frac{c}{2\pi f_0}. \quad (S9)$$

For example, the TiO$_2$ core radius is 11.6 μm for the resonant frequency $f_0 = 1.13$ THz.

Same as the previous work[4], coupling coefficients between resonators are calculated by COMSOL Multiphysics, deriving the exponential relation $\kappa/f_0 \sim 0.959 \cdot exp(-36 \cdot d/\lambda_0)$ for an excellent fit to $\kappa$ in the weak-coupling regime (here, $\kappa/f_0 < 1/40$, $d$: distance between resonators).